\documentclass[aps,pre,showpacs,amsmath,amsfonts,amssymb,superscriptaddress,twocolumn,nofootinbib]
{revtex4-1}

\usepackage[english]{babel}
\usepackage[pdftex]{graphicx}
\usepackage{tikz}
\usepackage{rotating}
\usepackage[pdftex,hypertexnames=true,linktocpage=true,breaklinks=true]{hyperref} 
\usepackage[hyphenbreaks]{breakurl}
\hypersetup{colorlinks=true,linkcolor=blue,anchorcolor=blue,citecolor=magenta,filecolor=blue,urlcolor=blue,bookmarksnumbered=false,pdfview=FitB}

\usepackage{amsthm}
\usepackage{framed}
\usepackage{amssymb}
\usepackage{amsmath,mathrsfs}
\usepackage{hhline}
\usepackage{ulem}
\usepackage{fancyhdr}
\usepackage{simplewick}

\everymath{\displaystyle}
\everymath{\displaystyle}

\newcommand{\affA}{Aix Marseille Univ, CNRS, CPT, Marseille, France}
\newcommand{\affB}{CNRS Centre de Physique Th\'eorique UMR7332,
13288 Marseille, France}
\newcommand{\affC}{Aix Marseille Univ, CNRS, INSERM, CIML, Marseille, France}

\newcommand{\affH}{Department of Oncology, 3-336, Cross Cancer Institute, 
Edmonton, AB, T6G 1Z2, Canada  }

\begin{document}

\title{Detection of long-range electrostatic interactions between charged molecules by means of Fluorescence Correlation Spectroscopy}

\author{Ilaria Nardecchia}
\email{i.nardecchia@gmail.com}
\affiliation{\affB}\affiliation{\affC}
\author{Mathias Lechelon}
\email{mathias.lechelon@gmail.com} 
\affiliation{\affA}\affiliation{\affB}\affiliation{\affC}
\author{Matteo Gori}
\email{gori@cpt.univ-mrs.fr}
\affiliation{\affA}\affiliation{\affB}
\author{Irene Donato}
\email{irene.donato@cpt.univ-mrs.fr} 
\affiliation{\affB}\
\author{Jordane Preto}
\affiliation{\affH}
\author{Elena Floriani}
\affiliation{\affA}\affiliation{\affB}
\email{floriani@cpt.univ-mrs.fr}
\author{Sebastien Jaeger}
\affiliation{\affC}
\author{Sebastien Mailfert}
\email{mailfert@ciml.univ-mrs.fr}
\affiliation{\affC}
\author{Didier Marguet}
\email{marguet@ciml.univ-mrs.fr}
\affiliation{\affC}
\author{Pierre Ferrier}
\email{ferrier@ciml.univ-mrs.fr}
\affiliation{\affC}
\author{Marco Pettini}
\email{pettini@cpt.univ-mrs.fr}
\affiliation{\affA}\affiliation{\affB}

\begin{abstract}
The present paper deals with an experimental feasibility study concerning the detection of long-range intermolecular interactions through molecular diffusion behavior in solution. This follows previous analyses, theoretical and numerical, where it was found that inter-biomolecular long-range force fields of electrodynamic origin could be detected through deviations from Brownian diffusion. The suggested experimental technique was Fluorescence Correlation Spectroscopy (FCS). By considering two oppositely charged molecular species in watery solution, that is, Lysozyme protein and a fluorescent dye molecule (Alexa488), the diffusion coefficient of the dye has been measured by means of the FCS technique at different values of the concentration of Lysozyme molecules, that is, at different average distances between the oppositely charged molecules. For the model considered long-range interactions are built-in as electrostatic forces, the action radius of which can be varied by changing the ionic strength of the solution. The experimental outcomes clearly prove the detectability of long-range intermolecular interactions by means of the FCS technique. Molecular Dynamics simulations provide a clear and unambiguous interpretation of the experimental results.
\end{abstract}
\date{\today}
\maketitle

\section{INTRODUCTION}
The present paper reports about the third and last step of a feasibility study concerning a possible strategy to detect long range electrodynamic forces acting among biomolecules. In living matter, resonant (thus selective) electrodynamic attractive interactions \cite{pre3} could be a relevant mechanism of molecular  recruitment at a distance, beyond all the well-known short-range forces (chemical, covalent bonding, H-bonding, Van der Waals). Unfortunately, because of technological limitations, an experimental proof or refutation of this possibility has been for a long time and is still sorely lacking. In our preliminary investigations in \cite{pre1} and \cite{pre2} we have put forward the idea that a possible experimental method to investigate whether these forces can be at work, when suitably activated, could come from the study of how the diffusion behavior of biomolecules in solution could change when their concentration is varied (that is, when the average intermolecular distance is varied).

The experimental technique envisaged in \cite{pre2} was Fluorescence Correlation Spectroscopy (FCS), a well established experimental technique \cite{Webb1974,Schwille2007,Elson2011}. In order to check whether the study of molecular diffusion through the FCS technique can be actually effective to detect intermolecular 
long-range interactions, in the present paper we have chosen to tackle a system of molecules which interact through {\it built-in} long range interactions, that is, through an electrostatic force field. Even if our ultimate goal is to detect electrodynamic intermolecular interactions, there are still several uncertainties concerning their activation and strength, therefore a preliminary assessment of the reliability of the experimental method - performed under perfectly controlled conditions - is timely. Thus we have now undertaken a successful experimental assessment of the method, a crucial step forward with respect to the two preceding works in Refs. \cite{pre1} and \cite{pre2}.
We chose a system 
where the interacting molecules were a protein, white egg Lysozyme, and an oppositely charged dye, Alexa Fluor 488. These molecules were solvated in pure water, thus in the absence of Debye screening, and in salted water to confirm that the concentration dependent attenuation of the self diffusion coefficient is due to the electrostatic interparticle interactions. Molecular Dynamics simulations have been also performed, and their results are in excellent agreement with the experimentally observed phenomenology. We conclude that the FCS technique is actually a viable experimental procedure for an assessment of the strength of long-range intermolecular interactions and thus, sooner or later, also of electrodynamic intermolecular interactions. The paper is organised as follows: in Section II the experimental results are reported and discussed, while in Sec. III  we report the outcomes of the Molecular Dynamics simulations of the experiments and we comment on the observed phenomenology. Section IV is devoted to some concluding remarks  about the results presented throughout the present paper. 
\section{EXPERIMENTAL RESULTS}
In the present Section we report about the effect of electrostatic long distance intermolecular interactions on the diffusion behavior of oppositely charged molecules. Molecular diffusion is detected  using Fluorescence Correlation Spectroscopy (FCS). As mentioned in the Introduction, the interacting molecules considered in the present study are Lysozyme, a small globular protein of   $14307$ Da \cite{Cantifield1963,Jolles1969} keeping a net positive charge for all pH values up to its isoelectric point, which is around pH $=11.35$  \cite{Wetter1951}, and Alexa Fluor 488 dye (hereinafter AF488) a very bright anionic fluorophore. FCS is a well established spectroscopic technique that enables a real time investigation of diffusion processes through a statistical analysis of the fluctuating fluorescence signal detected \cite{Webb1974,Schwille2007,Elson2011,RiglerElson2001,Webb2002}.   Self diffusion is affected by any interaction among the diffusing species, repulsive or attractive, that produce an attenuation depending on the interparticle interaction; the stronger the interaction the larger the deviation from Brownian diffusion \cite{pre2}. The experimentally accessible parameters  to implement this  study are the average intermolecular distance $\langle d \rangle$  and the ionic strength of the electrolytic solution used, as already discussed in a preceding paper of ours \cite{pre2}. The average intermolecular distance among molecules  changes with  their concentration as $\langle d \rangle= {C}^{-1/3}$, where $C$ is the total number of molecules per reaction volume. The  electrostatic interaction among the molecules in  electrolytic solution is described by the Debye-H\"{u}ckel potential \cite{Wright}:
\begin{equation} \label{Debye}
U_{\text{Debye}}({\bf r})= \dfrac{Z_1 Z_2 e^2}{\varepsilon_r |{\bf r}|} \cdot
\dfrac{\exp \left[-\frac{2R}{\lambda_D}\left( \frac{|\boldsymbol{r}|}{2R} - 1 \right)\right]}{\left(1+ R/\lambda_D \right)^2} \;,
\end{equation}
where $\lambda_D$ is the Debye screening length of the electrolytic solution, $R$ is the molecular radius, $e$ is the elementary charge and $\varepsilon_r$ is the static dielectric constant of the medium. For a monovalent electrolyte, like NaCl which has been used throughout this study, the Debye length in Eq.\eqref{Debye} reduces to:
\begin{equation}\label{Debyelength}
 \lambda _{D} = \sqrt{\frac{\varepsilon_r \varepsilon_0 k_B T}{2 N_A e^2 I}} \;.
\end{equation}
where $\varepsilon_0$ is the vacuum permittivity,  $k_B$ is the Boltzmann constant, $N_A$ is the Avogadro constant and $I$ is the ionic strength. Debye  screening - due to small ions freely moving in the environment -  is an essential feature of biological systems because it shortens the range of electrostatic interactions.   
In principle, counterions condensation effect can heavily affect the interaction among biomolecules so that "like likes like"
effects can take place \cite{Manning96,Gelbart2000}. Nevertheless for spherical macroions (as the proteins have been modeled
in numerical simulations) the counterions condensation phenomenon does not take place, and the Debye potential
properly accounts for the effect of counterions.
In the limiting case of  $\lambda_{D}\rightarrow +\infty$,  charged particles  in electrolytic solution are  submitted to a pure Coulomb potential given by:
\begin{equation}\label{Coul}
 U_{\text{Coul}}(\boldsymbol{r})= \frac{Z_1 Z_2 (e)^2}{\varepsilon_r |\boldsymbol{r}|} \;,
\end{equation}
that is, the Debye-H\"{u}ckel short-range potential turns to a long-range one. By long-range interaction  we mean an interaction potential falling off with the interparticle distance $r$ as $1/r^\nu $ with $\nu \leq d$, $d$ being the spatial dimension of the system.

The main outcome of FCS measurements on a solution of oppositely charged molecules is the average time $ \tau_D$ taken by a molecule of AF488 to cross the section of some observation volume in presence of different concentrations of Lysozyme. The measure of $ \tau_D$ gives access to physical quantities as the diffusion coefficient $D$ (Eq. \eqref{diffusion}) and the hydrodynamic radius $R_H$ (Eq.\eqref{stokes}). 
This study has been performed for different average values of the intermolecular interaction strength. The latter depends on  the average intermolecular distance $ \langle d \rangle$, and, possibly, on a variation of the Debye screening length. The average distance between any two interacting molecules is given by \cite{nota-concentr}
\begin{equation}
\langle d \rangle=(C_{AF488}+C_{Lys})^{-1/3} \;,
\label{Av-dist}
\end{equation}
where the concentration of the AF488 dye has been kept constant and equal to $1$ nM, while the Lysozyme concentration covered a range  between $0$ and $0.69$ mM ($9.86$ mg/ml).

\subsection{Autocorrelation and data treatment} 
\label{acf-data}

The autocorrelation function $G(\tau)$, originated by molecules interacting and diffusing in and out of the observation volume, is defined by 
\begin{equation}
G(\tau) = \dfrac{\langle \delta F(t) \delta F(t+\tau)\rangle}{\langle F(t)\rangle^2}
\label{ACFexp}
\end{equation}
 where $ \langle F(t)\rangle $ is the average intensity, $\delta F(t)$ the intensity of fluctuations, and the brackets mean ensemble average. The general procedure consists in fitting $G(\tau)$ with the appropriate mathematical model describing the characteristics of the system under study. The analytical form of the autocorrelation function  for a single molecular species, assuming a three-dimensional Gaussian profile of the excitation beam accounting for diffusion \cite{Webb1974} and a triplet state of the dye \cite{Rigler1995}, is obtained under the assumption that diffusion driven by random hits of water molecules  and protein-dye dynamics driven by electrostatic forces are independent processes:
\begin{equation}\label{ACF}
G(\tau)=1+\dfrac{1}{N} \,\dfrac{1+n_{T}\,\exp{\left(-\dfrac{\tau}{\tau_{T}}\right) } }{\left(1+\dfrac{\tau}{\tau_{D}}\right)\sqrt{1+s^{2}\dfrac{\tau}{\tau_{D}}}} \;.
\end{equation}
Here $N$ stands for the number of molecules in the FCS observation volume, $\tau_{D}$ is the diffusion time through this volume, $ \tau_{T} $ the  triplet lifetime, $n_{T}= Tr/(1-Tr)$, with $Tr$ the fraction of molecules in the triplet state.
The dimensionless parameter $s$, called structure parameter, describes the spatial properties of the detection volume. It is given by $s=\omega_{x,y} / \omega_{z}$, where the parameter $\omega_{z}$ is related to the length of the detection volume along the optical axis, and the radial waist $\omega_{x,y}$ is related to the radius of its orthogonal section.
 
The diffusion coefficient $D$ is expressed as a function of the radial waist $\omega_{x,y}$, and of the diffusion time $ \tau_{D}$ by:
 \begin{equation}\label{diffusion}
 D=\omega_{x,y}^{2}/4\tau_{D} \;,
 \end{equation}
and for isolated molecules following a  Brownian motion, the hydrodynamic radius $R_H$ may be computed using the Stokes$-$Einstein equation:
\begin{equation}\label{stokes}
R_{H}=\dfrac{k_B T}{6\pi \eta(T) D} \;,
\end{equation}
where $T$  is the absolute temperature, $k_B$ the Boltzmann constant, and $\eta$ the viscosity of the fluid. The viscosity of liquids is a decreasing function of temperature and is expressed empirically between $0^{\circ}$C and $370^{\circ}$C, with an error of 2.5 $\%$, by the expression \cite{viscosity}
\begin{equation}\label{eta}
\eta(T) = A \times 10^{B/(T - C)} \;.
\end{equation}
For water, the parameters $A, B$ and $C$ are equal to $2.414\times 10^{-5}$ Pa s, 247.8 K and 140 K, respectively.

\begin{figure}[h] \centering
\includegraphics [scale=0.6,keepaspectratio=true]{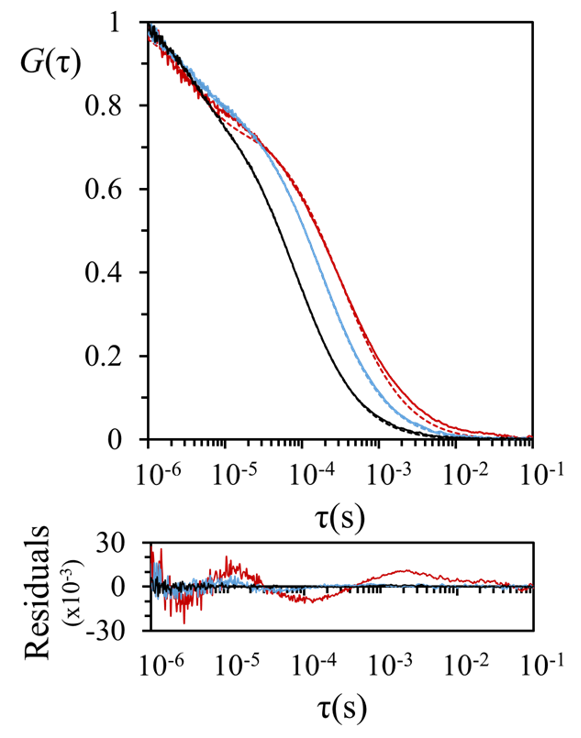}
\caption{(Color online) FCS measurements. Semilog plot of the normalized autocorrelation function of fluorescence fluctuations, defined in Eq.\protect{\eqref{ACF}}, obtained at $\langle d\rangle =240 \mathring{\text{A}}$ (red line), at $\langle d\rangle =920 \mathring{\text{A}}$ (blue line), and at $\langle d\rangle =4200  \mathring{\text{A}}$ (black line). Working temperature 30$^o$C.}
\label{ACFplot} 
\end{figure}

In \autoref{ACFplot} some typical outcomes of the FCS measurements are displayed. These are the  autocorrelation functions (ACFs) - defined in Eq.\eqref{ACFexp} - of fluorescence intensity fluctuations (for graphical reasons the normalized versions are displayed). 
To assess afterpulsing artefacts on the ACFs,  Fluorescence Cross Correlation Spectroscopy (FCCS) measurements  have been also performed, and the results so obtained  are in perfect agreement with those found with FCS (see Appendix).

Then the experimental ACFs are fitted by means of the analytic expression in Eq.\eqref{ACF} where we used $s=\omega_{x,y} / \omega_{z}=0.2$ and $ \tau_{T} $ was left free. Out of these measurements and fittings one obtains the diffusion times $ \tau_D$ at different values of $\langle d\rangle$, and hence, according to Eq.\eqref{diffusion},  the values of the diffusion coefficient $D$ of the AF488 molecules can be worked out after having performed an accurate measurement of the waist size $\omega_{x,y}$ by means of Rhodamine Rh6G used as  a diffusion standard (details are given in  Section \ref{Materials and Methods}).

When the Lysozyme concentration is zero the solution contains only 1nM of AF488 corresponding to an average intermolecular distance of ${11841.8 \mathring{\text{A}}}$.  The diffusion coefficient of the dye in the absence of Lysozyme is used  as the infinite dilution value $D_0 $ of AF488. Then the average protein-dye  distance is varied by varying the Lysozyme concentration.
The fitted values of the parameters for solutions of variable concentrations of Lysozyme and 1nM of AF488 are reported in the Table shown in \autoref{Tableau}. The number of fluorescent dye molecules $N$ is observed to decrease at increasing Lysozyme concentration as a consequence of a quenching phenomenon already reported in the literature and briefly discussed in Appendix A.

As several experiments have been performed on different days - so that the outcomes of the measures can be affected by even minor modifications of the FCS and FCCS setups -  the values of the measured diffusion coefficient $D$ are normalized to the infinite dilution value $D_0$, which is determined anew each time a new experiment is performed, and which corresponds to Brownian diffusion of the dye molecules. 

\begin{figure}[h] \centering
\includegraphics [width=90mm]{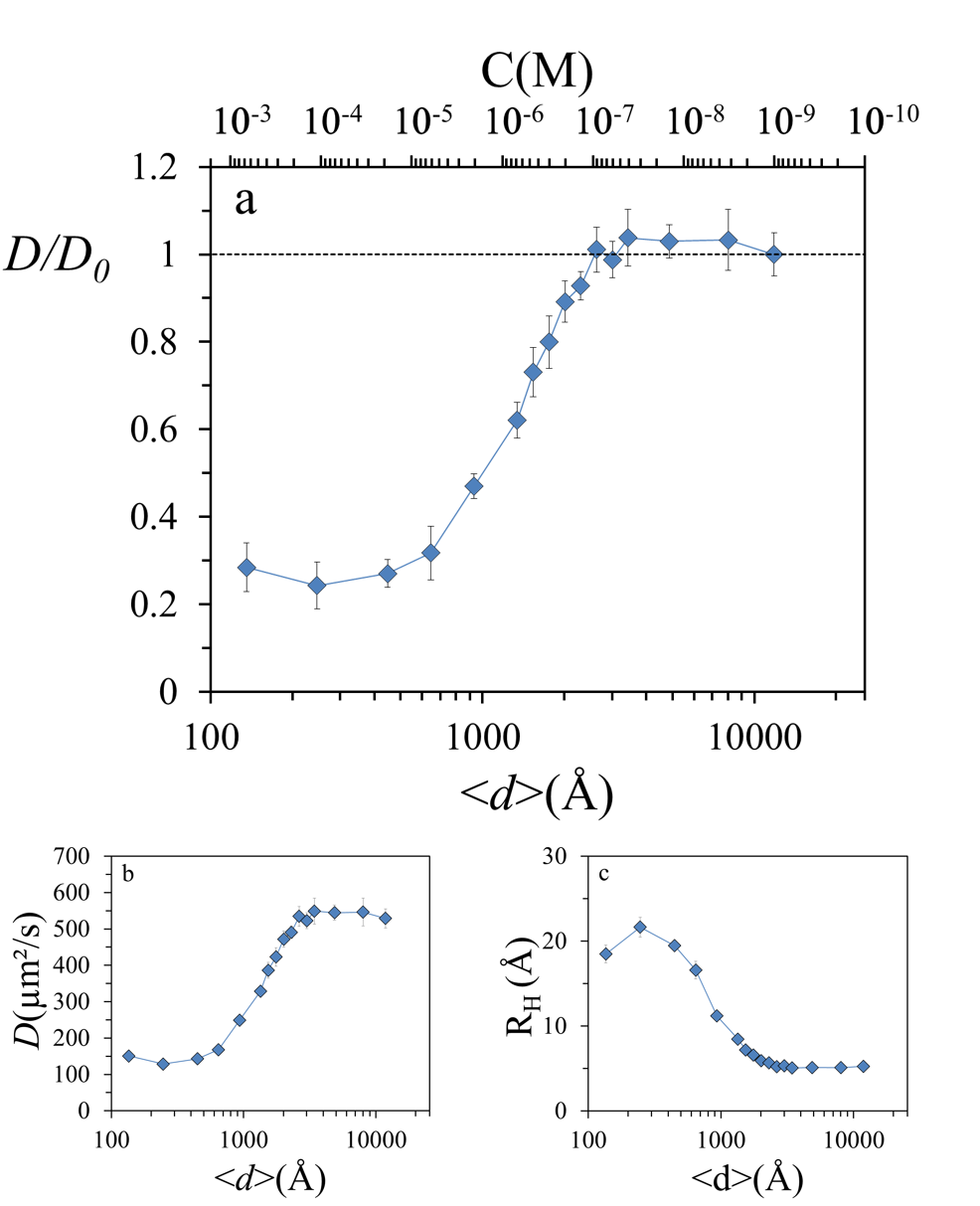}
\caption{(Color online) Semilog plot of the normalized diffusion coefficients $D/D_0$ of AF488 as a function of the distance in  $\mathring{\text{A}}{}$ between proteins and dyes (a); semilog plot of the diffusion coefficient $D$ for a single experiment with 0mM of NaCl in solution (b); semilog plot of hydrodynamic radius (c) of AF488 (1nM) versus the  average distance between all the molecules in pure water.}
\label{D-D0_D_Rh} 
\end{figure}

In \autoref{D-D0_D_Rh}(a) we can observe that at low concentrations of Lysozyme, corresponding to an average interparticle distance larger than approximately $2500 \mathring{\text{A}}$, the diffusion of the dye molecules is Brownian, that is, $D/D_0 \simeq 1$, where $D/D_0 = \tau_{D_0}/\tau_D$ (see Eq.\eqref{diffusion}). By increasing the Lysozyme concentration the normalised diffusion coefficient is observed to markedly drop, and this is attributed to the attractive electrostatic interaction among the dye and protein molecules, as is qualitatively discussed below, and quantitatively discussed in Section \ref{BasicEq}. 

The larger the concentration of protein molecules the stronger the electrostatic attraction they exert upon the dye molecules. The dye diffusion can be slowed down - below the Brownian diffusion regime - through two mechanisms: on the one side, being attracted in every direction, the dye molecules undergo a sort of dynamical "frustration''  \cite{pre2}; on the other side, the protein and the dye molecules can form temporary/"flickering" bound states. 

In \autoref{D-D0_D_Rh}(b) the experimental outcomes obtained for the non-normalised $D$  are plotted as a function of the average intermolecular distance. The knowledge of the diffusion coefficient $D$ allows to estimate the equivalent hydrodynamic radius $R_H$ through the Stokes-Einstein equation (Eq.\eqref{stokes}); the variation of $R_H$ as a function of $\langle d\rangle$ is reported in \autoref{D-D0_D_Rh}(c).

At infinite dilution, the  diffusion coefficient $D_0$ of AF488 in water, and its corresponding hydrodynamic radius, are found to be equal to $532 \pm 23.5\ \mu$m$^2$/s and $5.2 \pm 0.2\ \mathring{\text{A}}$, respectively.

\begin{figure}[h] \centering
\includegraphics [width=95mm]{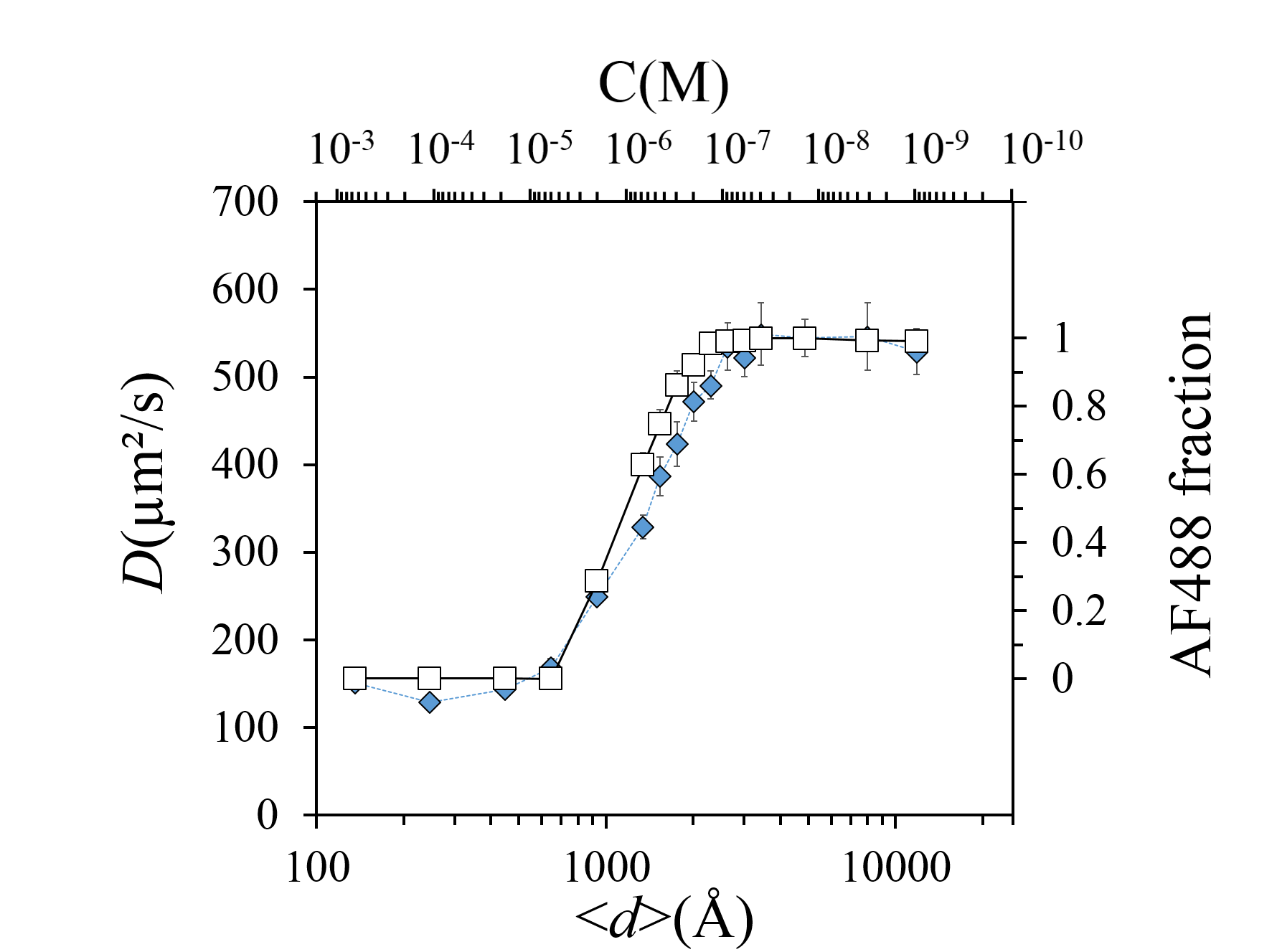}
\caption{(Color online) The diffusion coefficient (blue diamonds) and the fraction of free AF488 molecules (white squares) are compared. In the two-species fitting to compute the fraction of free dye molecules the following parameters have been kept fixed: $\tau_{AF488} = $78.8$\mu$s,  $\tau_{Lys} = $210$\mu$s, triplet time of AF488 3.4$\mu$s, and triplet time of Lysozyme 3.34$\mu$s.}
\label{dye-Lys} 
\end{figure}

\begin{figure}[h] \centering
\includegraphics [width=90mm]{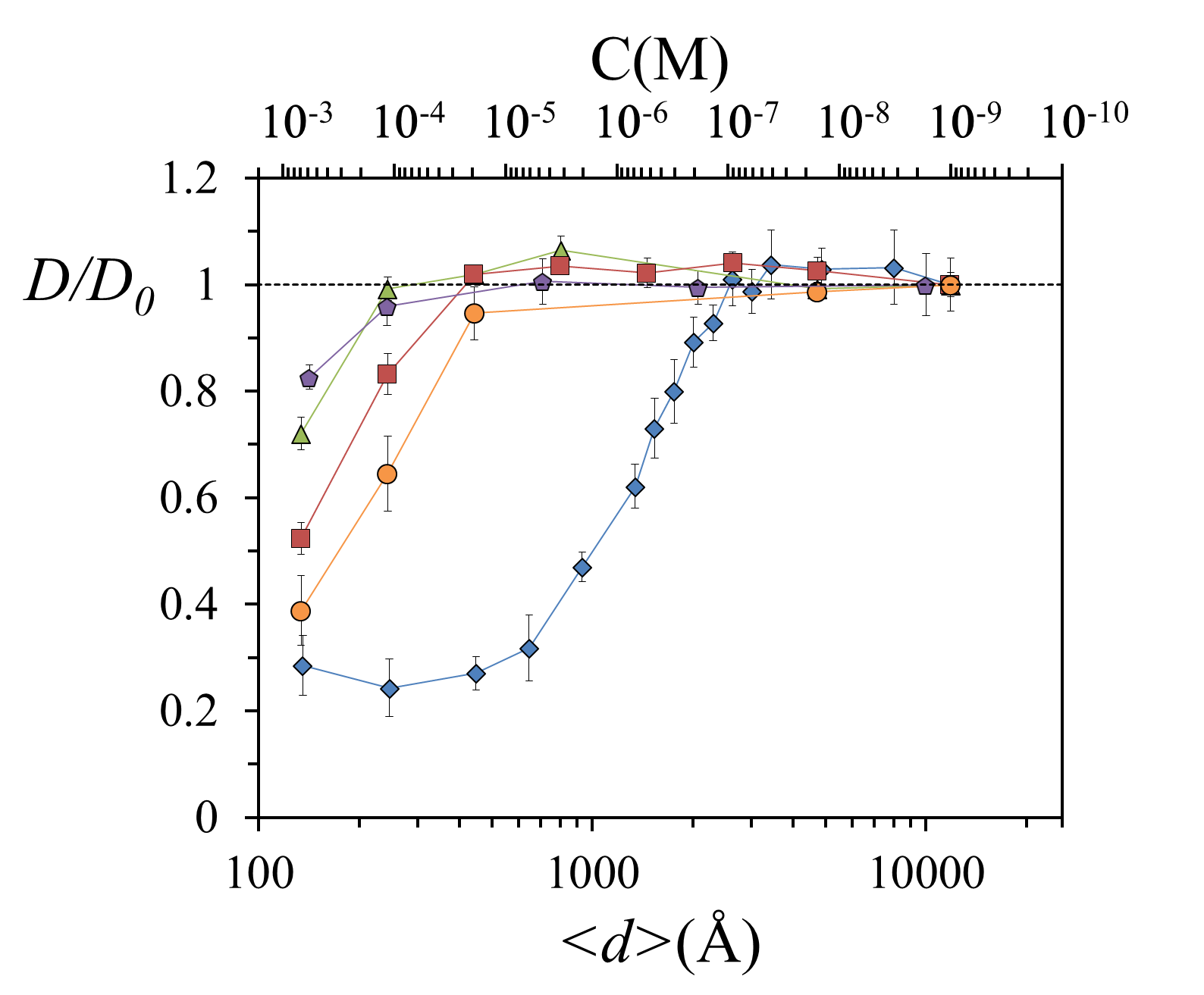}
\caption{(Color online) Semilog plot of the normalized diffusion coefficients $D/D_0$ at different concentrations of NaCl in solution: 0mM (blue diamonds), 20mM (orange circles), 50mM (red squares), 100mM (green triangles) and 150mM (purple rhombs).}
\label{D-D0_NaCl} 
\end{figure}

Let us remark that the patterns of the ACFs reported in \autoref{ACFplot} and  \autoref{ACFplotbis} are well fitted by the single species function in Eq.\eqref{ACF} even though a-priori a two-species ACF \cite{nota} could better take into account the possible presence of at least two subpopulations, one of dye molecules temporarily bounded to proteins, and the other of freely moving dye molecules. However, fitting the ACFs by taking as free parameters both the diffusion time of the AF488 molecules and of the Lysozyme molecules leads to a poor quality results. 

To the contrary, an interesting result is found by performing a two-species fitting, where both the diffusion times of AF488 and of Lysozyme are kept fixed, to work out the relative population of free AF488 dye versus temporary bound AF488-Lysozyme states. The outcome is reported in \autoref{dye-Lys} in the form of the fraction of freely moving AF488 molecules with respect to the total population of AF488 molecules, and given as a function of both the average interparticle distance $\langle d\rangle$, given by Eq. \eqref{Av-dist}, and particle concentration. It appears evident that when the diffusion of the AF488 molecules is Brownian the fraction of freely diffusing molecules is equal to 1, then this fraction drops parallely to the drop of the diffusion coefficient until almost all the AF488 molecules appear bounded to the Lysozyme molecules. 
This cross checks with the  observation that the diffusion coefficient $D$ measured at the lowest values of the intermolecular distances has to correspond to the condition where all the molecules of AF488 are bounded to the Lysozyme molecules for most of the time, thus in this case $D$ must approximately equal the value of the Lysozyme diffusion coefficient (apart from a small difference due to a slightly modified Stokes radius).
Therefore, the diffusion coefficient of Lysozyme molecules chemically labelled with AF488 has been measured. The value obtained is $D_{Lys}=166.55 \pm 1 \mu$m$^2$/s which is in fairly good agreement with that one corresponding to the lowest intermolecular distance in \autoref{D-D0_D_Rh}(b). 

As a control, we use the white egg Lysozyme hydrodynamic radius $R_H= 20.5 \mathring{\text{A}}$ reported in Ref.\cite{Wilkins1999}, and resorting to Eq.\eqref{stokes} relating $R_H$ to the diffusion coefficient together with Eq. \eqref{eta} for the water viscosity at our working temperature of $30^{\circ}$C, one finds $D_{Lys}=138 \mu$m$^2$/s which is in fairly good agreement with our above reported value, considering that the calibration of our device is based on the outcomes of Ref. \cite{Schwille2008}.

The temporary nature of these bound states is reasonably surmised because the random hits of water molecules, which drive molecular diffusion, are also continuously destroying the bound states. To support these qualitative arguments we can proceed by estimating and comparing some characteristic energy and time scales. A characteristic interaction length $R_{el}$ is obtained by equating the electrostatic interaction energy with thermal energy giving
\begin{equation}
R_{el}=\dfrac{|Z_{Lys}Z_{dye}| e^2}{4 \pi \varepsilon_r k_{B} T}
\end{equation}
taking $Z_{Lys}=10,Z_{dye}=-2$, $e$ the electron charge, and the dielectric constant of water $\varepsilon_{r}\sim 76$ at 
$T=303 K$ (30$^{\circ} C$), we have
$R_{el}\sim 145 \mathring{A}$
so that the mean time taken by a dye molecule to move on such a distance is 
\begin{equation}
\tau_{0}=\dfrac{R_{el}^2}{6 D_{0}}\sim 66\,\, ns 
\end{equation}
having assumed $D_{0}\simeq 532 \mu$m$^2 s^{-1}$ (see below). The largest value of the electrostatic interaction energy between AF488 and Lysozyme is estimated as
\begin{equation}
E_{max}=\dfrac{Z_{Al}Z_{Lys} e^2}{4\pi \varepsilon_{r} R_{min}}\sim 150 \,\, meV
\end{equation}
where $R_{min}=R_{dye}+R_{Lys}=25 \mathring{A}$ is the minimum distance between AF488 and Lysozyme, and since at 30$^{\circ} C$ $k_BT\simeq $26 meV, 
we compute the Kramers  escape time from a bound state as the time for an AF488 molecule to reach a distance $R_{el}$, where the electrostatic interactions are equal to the thermal energy $k_BT$, that is
\begin{equation}
\tau_{Kr}=\tau_{0}\exp\left[\dfrac{\Delta E}{k_{B}T}\right]\simeq 9\, \mu s
\end{equation}
where $\Delta E= E_{max} - k_BT$; this gives an order of magnitude of the average trapping time of a molecule of AF488 by a Lysozyme molecule.
For instance, at $\langle d \rangle=1000 \mathring{A}$  the free diffusion time of an AF488 molecule is
\begin{equation}
\tau_{free}(1000 \mathring{A})=\dfrac{(\langle d \rangle-2R_{el})^2}{6 D_{0}}\simeq 1.5 \mu s
\end{equation}
which is shorter than the average trapping time, as a consequence for this value of  $\langle d \rangle$ the AF488 molecules are  most of the time, but not permanently,  trapped, and this is consistent with a small value of $D/D_0$. To the contrary,  for $\langle d \rangle=2500 \mathring{A}$
the free diffusion time is approximately given  by
\begin{equation}
\tau_{free}(2500 \mathring{A})\simeq 17\,\mu s
\end{equation}
which is longer than the trapping time. Of course these are somewhat crude estimates but provide reasonable orders of magnitude and thus useful heuristic information.
Finally, the existence also of temporary states of more than one dye molecule bounded to a single protein molecule is not excluded, but such a possibility is implicitly taken into account by the numerical simulations reported in Section \ref{BasicEq}.  

The above given natural explanation of the result displayed in \autoref{D-D0_D_Rh}(a) can be further and nicely confirmed by acting on the effective range of the the intermolecular interaction potential according to  Eq.\eqref{Debye}. The range of the potential is controlled by the Debye screening length $\lambda_D$, which depends on the concentration of freely moving ions in the electrolytic solution. This is practically realised by  adding to the watery solution of proteins and dyes different concentrations  of sodium chloride. 
In order to change the action range of electrostatic interactions we chose five different  NaCl concentrations: $0, 20, 50, 100, 150$ mM. The $0$ mM concentration  of NaCl implies that the  molecules are solvated in pure water and submitted to a pure Coulombic potential (Eq.\eqref{Coul}), while the additions of salt in solution screens the electrostatic interaction  between charged molecules (Eq.\eqref{Debye}). The Debye screening lengths  for NaCl salt contents of $20, 50, 100$ and $150$ mM,  are equal to $21.4 \mathring{\text{A}}{}$, $13.6 \mathring{\text{A}} {}$,  $9.6 \mathring{\text{A}} {}$ and $7.8 \mathring{\text{A}} {}$ (Eq.\eqref{Debyelength} for a temperature of $30^{\circ}$C), respectively. 

The effect of this action on $\lambda_D$ is shown in \autoref{D-D0_NaCl}. The different patterns of $D(\langle d \rangle)$ are consistently showing that the higher the ionic strength (that is the shorter $\lambda_D$) the shorter the distance $\langle d \rangle$ at which $D$ deviates from a Brownian value.

\begin{figure}[h] \centering
\includegraphics [width=75mm]{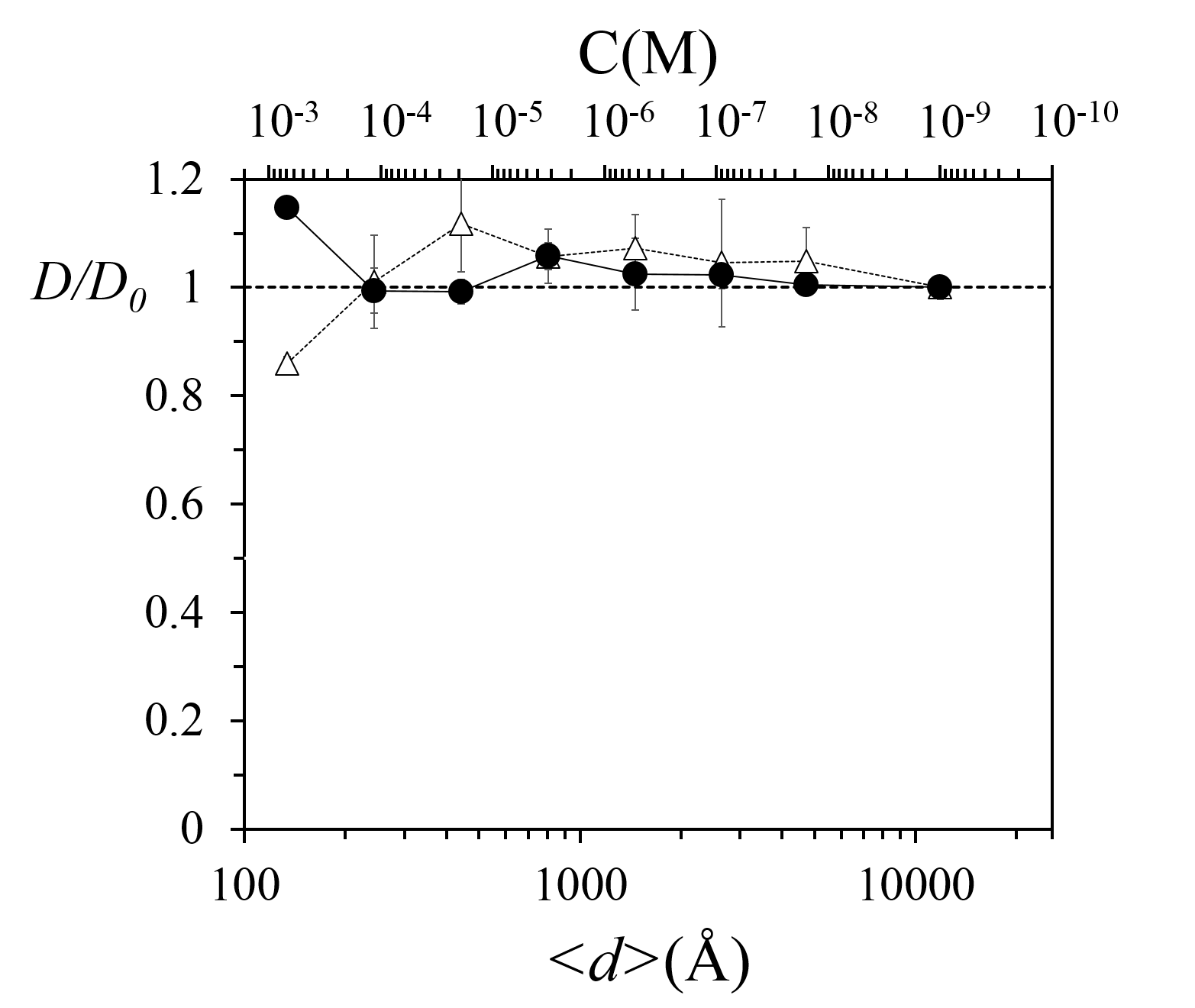}
\caption{(Color online) Normalised diffusion coefficients at isoelectric points of Myoglobin at pH=7 (full circles), and of Lysozyme at pH=11 (open triangles). Electrostatic interactions are switched off.}
\label{pH11} 
\end{figure}

In order to further cross check that the observed lowering of the diffusion coefficient of the dye is due to its electrostatic interaction with the protein molecules, we have replaced Lysozyme (molecular weight = 14.3 kDa) with Myoglobin (molecular weight = 16.7 kDa). We chose Myoglobin because its isoelectric point happens to occur at pH=7, the pH value of our measurements, thus a mixture of AF488 and Myoglobin at pH=7 in pure water is expected to always give Brownian diffusion. And this is actually the case, as is shown by \autoref{pH11}. Then we also considered a solution of Lysozyme and AF488 at pH=11, the isoelectric point of Lysozyme. Also in this case no trace is left of the pattern of the diffusion coefficient reported when both the molecular species are charged (though in this case the buffer keeping at 11 the pH of the solution has a non-vanishing ionic strength). 
The fit of our ACFs using a $3D$ anomalous diffusion model (see Appendix \ref{Ap}) has not highlight the existence of  molecular crowding effects in our system.
The last cross check  has been the fit of our ACFs using a $3D$ anomalous diffusion model  to study how molecular crowding effects were affecting our system. The results clearly show that crowding effects are negligible (see Appendix).

\section{Materials and Methods}\label{Materials and Methods}

\subsection{Materials}\label{Materials}

Chicken egg white Lysozyme was purchased from Sigma (L6876) (St. Louis, MO); it was used without further purification and was either dissolved directly into bidistilled water or diluted in series. Even though dissolving a protein in pure water can strongly modify its 3D structure, in the present context this does not affect the phenomenon in which we are interested.

AF488 5-TFP (Alexa Fluor $488$ Carboxylic Acid, $2,3,5,6-$Tetrafluorophenyl Ester), $5$-isomer (A$30005$) was purchased from Molecular Probes Invitrogen.

The dye has excitation/emission of 495/515 nm and a molar extinction coefficient of $ \varepsilon_{495}  = 71,000$ M cm$^{-1}$. Both the protein and dye concentrations have been determined measuring their absorbance with a Nanodrop 1000 Spectrophotometer (ThermoScientific), at $280$ nm with a molar extinction coefficient of $36000$ M cm$^{-1}$ and at 495 nm with a molar extinction coefficient of $71000$ M cm$^{-1}$, respectively. To change the ionic strength of the solution, sodium chloride (NaCl) has been diluted in bidistilled water getting molar concentrations of 20mM, 50mM, 100mM and 150mM. Then these solutions  are used to dilute the protein and the dye. The above mentioned chemically labelled Lysozyme molecules have been obtained by homemade labelling using 2 mg/mL of proteins with 0.7 times of A488 5-TFP in sodium bicarbonate at a pH $\approx$ 8.5, during one hour. This has been done in order to compare the results of AF488 diffusion time in our experiments with the Lysozyme diffusion time. Unconjugated dye was removed using a PD-mini-trap G25 (GE Health Care) according to the instructions of the manufacturer, using gravity protocol, and the degree of labelling was determined spectroscopically.
As a container for the solutions, 8-wells Labtek chambered coverglass has been used.

\subsection{FCS measurements}
 FCS measurements were performed on a custom-made apparatus based on an Axiovert 200 M microscope (Zeiss, Germany) with an excitation 488 nm Ar$^{+}$-ion laser beam focused through a Zeiss water immersion Apochromat 40X/1.2 numerical aperture objective. The fluorescence was collected by the same objective, separated from the excitation light using a dichroic mirror, and then delivered to an avalanche photodiode (SCPM AQR-13, Perkin Elmer) through 545/20 nm bandpass filter. A 50 $\mu$m diameter confocal pinhole reduced the out-of-focus fluorescence. The system has been switched on  around 60 min prior to the measurements to allow for stabilization of all components. The laser waist $ \omega_{x,y}$ was set by selecting with a diaphragm the lateral extension of the laser beam falling onto the back-aperture of the microscope objective \cite{Marguet2005} and was then estimated using  the  well known diffusion of Rhodamine $6G$ in water \cite{Schwille2008} $\omega_{x,y}=\sqrt{4 D \tau_D}$. The literature diffusion coefficient for Rh6G \cite{Rhodam} ($D_{Rh6G,22.5^{\circ}C}= 426 \mu$m$^2/$s) has been corrected to account for the experimental temperature ($30^{\circ}$C) and the viscosity of water at the same temperature (Eqs.\eqref{stokes} and \eqref{eta}), giving $D_{Rh6G,30^{\circ}C} = 517 \mu$m$^2/$s. For Rhodamine 6G sample we used a power of $300 \mu$W at the back-aperture objective, while for all the other samples, to avoid any saturation effect that would affect the shape of the excitation volume, we used a power of $100 \mu$W.

The measurements have been performed in 8 well chamber slide using a volume of 400 $\mu$L. In order to check whether some artifacts could be caused by the sticking of either the dye or the protein molecules to the well walls, some tests have been performed with coated labteks. The coating was performed overnight, at room temperature, by putting 700$\mu$L of BSA in PBS at a concentration of 10 mg/ml in each well of the labtek. Then the labtek has been gently rinsed two times and dried.
Tests performed with coated labtek have not shown differences (data not shown). The measurement time was of 200 s divided into 10 s runs. The experiments were performed at a temperature of 30$^\circ$C and the  samples  were  thermostated in the microscope system at least during two hours. Each measurement was repeated at least three times successively in different labtek wells. The AF488 concentration was fixed at 1 nM in all samples. The diffusion coefficients were determined by measuring the apparent diffusion time of the fluorescent molecules through a confocal volume, always using the formula $D=\omega_{x,y}^{2}/4\tau_{D}$. The normalization of the diffusion coefficients was done by taking for $D_0$ the value of the diffusion coefficient of the AF488 obtained at 1 nM. Each reported value of $D$ and its corresponding error bar are the outcomes  of 20 measurements performed on four different samples and repeated for three independent experiments. 

Autocorrelation function calculations and fits were performed using the analytical expression given by Eq.\eqref{ACF}. The ACF data were fitted to estimate the parameters $N$, $\tau_{D}$, $\tau_{T}$. Then, the resulting value of $N$ and the corresponding count rate, were used to determine the brightness per molecule. The accuracy of the fit for each data set was assessed through the value of the $\chi^2$  parameter and by inspection of the residuals, which were to be distributed uniformly around zero.

\subsection{FCCS measurements}\label{fccsmeas}
FCCS measurements were performed on a commercial FCS setup (ALBA FCSTM, from ISS Inc., Champaign, America) with two excitation picosecond/CW diode lasers operation at 488 and 640 nm (BDL-488-SMN, Becker and Hickl, Germany) with a repetition rate of 80 MHz, focused through a water immersion objective (CFI Apo Lambda S 40X/1.25 WI, Nikon). The fluorescence was collected by the same objective, splitted into two detection paths by a 50/50 beam splitter (Chroma 21000) and filtered by two Emission filters (525/40 nm band pass, Semrock FF02-525/40  and 675/67 nm band pass, Semrock FF02-675/67 for the green and red channels, respectively) and detected by two avalanche photodiodes (SPCM AQR-13 and SPCM ARQ-15, Perkin Elmer / Excelitas). 
\begin{figure}[h!] \centering
\includegraphics [width=100mm]{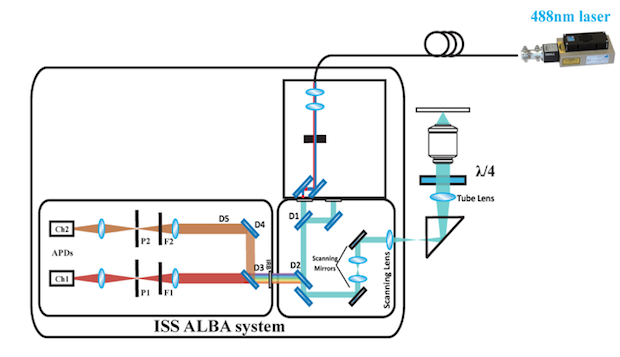}
\caption{(Color online) Schematic diagram of the FCCS apparatus. The excitation beam starts from the diode laser (on the top right of the image), goes through an optic fiber, through a set of lenses and a dichroic mirror (on the middle of the image), and reaches the sample placed in a confocal microscope through the objective (on the right). The emitted fluorescence beam goes back through the objective, is reflected on the dichroic mirror, then is splitted into two beams that are finally collected by two detectors (on the left). Other details in  \protect{\autoref{fccsmeas}}. }
\label{fccs} 
\end{figure}

\section{Numerical simulations}\label{BasicEq}
We consider a system composed of two different molecular species $A$ and $B$, modelled as spherical Brownian particles of radii $R_A$, $R_B$, and of net number of electric charges $Z_A$, $Z_B$, respectively. We refer to the number of each type of particle as $N_A$ and $N_B$, respectively, and to $N=N_A+N_B$ as the total number of particles. Such molecules move in a fluid of viscosity $\eta$ at a fixed temperature $T$. The friction exerted by the fluid environment on the particles is described by the Stokes' law $\gamma_{i}=6\pi\eta R_i$. Particles interact through pairwise potentials $U(r)$ which depend on the distance $r$ between their centers. For the mutual distance between the $i_k$ particle of type $k$ and the $j_l$ particle of type $l$  we introduce the notation:
\begin{equation}
r_{i_k j_l}=\left|\mathbf{r}_{i_k}-\mathbf{r}_{j_l} \right| \;,
\end{equation}
where $\boldsymbol{r}_{i}$ is the coordinate of the center of $i$-th particle so that $U_{kl}(r_{i_k j_l})$ represents the total interaction energy between these particles. We work under the assumption that the dynamics of the system is given by $N$ coupled Langevin equations in the so called overdamped limit, where inertial effects are neglected \cite{pre2,Gardiner2009}:
\begin{eqnarray}
&\dfrac{d \boldsymbol{r}_{i_A}}{dt} = -\dfrac{1}{\gamma_A}\Biggr[ \sum_{\substack{j_A=1\\ j_A \neq i_A}}^{N_A}\boldsymbol{\nabla}_{\boldsymbol{r}_{i_A}} U_{AA}\left(r_{i_A j_A}\right)+\nonumber\\
&-\sum_{j_B=N_A+1}^{N_A+N_B}\boldsymbol{\nabla}_{\boldsymbol{r}_{i_A}} U_{AB}\left(r_{i_A j_B}\right)
\Biggr]+\sqrt{\dfrac{2 k_B T}{\gamma_A}} \boldsymbol{\xi}_{i_A}(t)\nonumber\\
\end{eqnarray}
\begin{eqnarray}
&\dfrac{d \boldsymbol{r}_{i_B}}{dt}=-\dfrac{1}{\gamma_B}\Biggr[\sum\limits_{\substack{j_B=N_A+1\\j_B\neq i_B}}^{N_A+N_B} \boldsymbol{\nabla}_{\boldsymbol{r}_{i_B}}U_{BB}\left(r_{i_B j_B}\right)+\nonumber\\
&-\sum\limits_{j_A=1}^{N_A} \boldsymbol{\nabla}_{\boldsymbol{r}_{i_B}} U_{AB}\left(r_{i_B j_A}\right)\Biggr]+\sqrt{\dfrac{2 k_B T}{\gamma_B}} \boldsymbol{\xi}_{i_B}(t)\\
\nonumber\\
&\qquad i_A=1,...,N_A \qquad i_B=N_A+1,...,N_A+N_B \nonumber
\label{lang2.}
\end{eqnarray}
where $k_B$ is the Boltzmann constant, and $\boldsymbol{\xi}(t)=(\boldsymbol\xi_1,...,\boldsymbol\xi_N)$ stands for a $3N$-dimensional random process modelling the fluctuating force due to the
collisions with water molecules, which is assumed to satisfy the following relations:
\begin{equation}
\begin{cases}
\langle \xi^{\alpha}(t)\rangle_{\xi}=0\\
\\
\langle \xi^{\alpha}_{i}(t) \xi^{\beta}_{k}(t')\rangle_{\xi} =
\delta^{\alpha\beta} \delta_{ik} \delta(t-t') \\
\end{cases}
\label{whitenoise}
\end{equation}

\subsection{Model potential}\label{AnalyticalPotential}
The interactions among the molecules are linear combinations of pairwise potentials regularized as follows:
\begin{equation}
\label{eq:InterPotential}
U_{kl}(r_{i_{k}j_{l}})=
\begin{cases}
U_{SC}(r_{i_{k}j_{l}})  \qquad &r_{i_{k}j_{l}}\leq 1.01\, R_{kl}
\\
U_{ElStat}(r_{i_{k}j_{l}})  \qquad &r_{i_{k}j_{l}}>1.01\, R_{kl}
\end{cases}
\end{equation}
where $R_{kl}=R_k+R_l$ is the sum of the two molecular radii, $U_{SC}(r_{i_{k}j_{l}})$ is a soft-core potential and $U_{ElStat}(r_{i_{k}j_{l}})$ is the Coulomb electrostatic potential. The choice of a soft-core potential is related to the fact that a small ficticious compenetration among the interacting molecules is allowed in numerical simulations for computational reasons (that is, to avoid the need of very short integration time steps). The soft-core potential has the form:
\begin{equation} 
U_{SC}(r_{i_{k}j_{l}})=A_{SC_{kl}}\exp{\left[-\dfrac{r_{i_{k}j_{l}}}{R_{kl}}+1\right]} \;.
\end{equation}
The parameter related with the potential strength $A_{SC_{kl}}$ has been chosen such that:
\begin{equation}
\Delta r_{SCi_{k}}(0.95\, R_{kl}) +\Delta r_{SCj_{l}}(0.95 \,R_{kl} )=0.05 \, R_{kl} \;,
\end{equation}
where $\Delta r_{SCi_{k}}$ is the drift of the particle $i_{k}$ due to the soft-core potential in a discrete time interval $\Delta t$:
\begin{equation}
\Delta r_{SCi_{k}}(r_{i_{k}j_{l}})=\dfrac{\Delta t}{\gamma_k}\left|\dfrac{\mathrm{d}U_{SC}(r_{i_{k}j_{l}})}{\mathrm{d}r}\right| \;.
\end{equation}
This yields the following expression for $A_{SC_{kl}}$:
\begin{equation}
A_{SC_{kl}}=0.05\, \exp\left[-0.05\right] \,\dfrac{R_{kl}^2}{\Delta t}\left(\dfrac{1}{\gamma_k}+\dfrac{1}{\gamma_l}\right)^{-1} \;.
\end{equation}
The electrostatic Coulomb potential $U_{ElStat}(r_{i_{k}j_{l}})$, describing the experimental condition where no salt is dissolved in solution, is:
\begin{equation}\label{Coulomb}
 U_{\text{Coul}}(r_{i_{k}j_{l}})= \dfrac{Z_k\, Z_l \, e^2}{\varepsilon \, r_{i_{k}j_{l}}} \;,
\end{equation}
where $e$ is the elementary charge and $\varepsilon$ is the electric permittivity of the medium, for which the static value at room temperature is $\varepsilon=\varepsilon_{water}\simeq 80$.

\subsection{Numerical algorithm}\label{subsec:numAlg}
We have numerically studied systems of two populations of molecules confined in a cubic volume of size $L$. The number of particles for each type is fixed: $N_A$ is the number of $A$-type particles (Lysozyme molecules), and $N_B$ is the number of $B$-type particles (AF488 molecules). To avoid spurious boundary effects, periodic boundary conditions (PBC) have been assumed, which is equivalent to  the existence of an infinite number of images/replicas throughout the space. In order to study diffusion at different concentrations, the numbers of molecules $N_A$ and $N_B$ are kept fixed, and the average intermolecular distance $\langle d \rangle$ among the molecules of type $A$ and $B$ is then controlled according to the relation
\begin{equation}
L=\sqrt[3]{N_A + N_B}\,\langle d \rangle \;.
\end{equation}

We remark that such a choice is not entirely equivalent to the experimental situation described in the previous paragraphs where the dye (AF488) concentration was fixed; in fact, in molecular dynamics simulations (MDS) both the concentration of Lysozyme and dye change with $\langle d \rangle$, the ratio of concentrations being constant. This choice is justified by the fact that in real experiments $N_A/N_B$ varies in a range $[1 - 5\times10^{5}]$; fixing $N_B=50$, the experimental situation would correspond to taking $N_A$ in a range $[1 - 2.5\times\,10^{6}]$, which is very highly demanding for computation. In MDS the ratio $N_A/N_B$ has been chosen as large as possible ($N_A/N_B= 10$ in our case) to avoid that dye molecules (AF488) dynamics could significantly affect the biomolecules (Lysozyme) dynamics.

In the presence of long-range interactions and PBC, each molecule contained in the previously mentioned box interacts with all the molecules contained in the above mentioned images/replicas, that is, the pairwise potential
\begin{equation}
U_{kl}(r_{i_{k}j_{l}})=U(|\boldsymbol r_{i_{k}}-\boldsymbol r_{j_{l}}|)
\end{equation}
in Eq. \eqref{lang2.} has to be replaced by an effective potential $U^{\mathrm{eff}}_{kl}(r_{i_{k}j_{l}})$ of the form:
\begin{equation}
U^{\mathrm{eff}}_{kl}(r_{i_{k}j_{l}})=\sum_{\boldsymbol{n}\in\mathbb{Z}^3}U(|\boldsymbol{r}_{i_k}-\boldsymbol{r}_{j_l}+\boldsymbol n L|) \;,
\label{effpotPBC}
\end{equation}
where $\mathbb{Z}^3$ is the space of $3$-dimensional integer vectors. 

It is clear that short and long-range interactions (in the sense specified in the Introduction) have to be managed in two different ways. For short range interactions it is always possible to define a cutoff length scale $\lambda_{\text{cut}}$ such that the effects of the interactions beyond this distance are negligible. In the systems we have studied by means of numerical simulations, the Debye electrostatic potential is a short range potential with a cutoff scale of the order of some units of the Debye length $\lambda_D$. 
For long-range interactions as the Coulomb potential Eq.\eqref{Coulomb}, it is not possible to define a cutoff length scale $\lambda_{\text{cut}}$ so that, in principle, an infinite sum should be considered. A classical way to account for long-range interactions resorts to the so called Ewald summation \cite{Allen1989}. In the following Section we describe a more recent and practical method - replacing Ewald's one - known as Isotropic Periodic Sum (IPS). 
The equations of motion \eqref{lang2.} were numerically solved using a second order Euler-Heun algorithm \cite{Burrage2007},  that is, a predictor-corrector scheme. 

The initial position of each particle is randomly assigned at $t_0$ using a uniform probability distribution in a cubic box of edge $L$.
\subsubsection*{IPS correction to long-range potentials} 
Because of the long-range nature of the Coulomb potential described by Eq.\eqref{Coulomb}, the force acting on each particle is given by the sum of the forces exerted by all the particles in the box and by the particles belonging to the images. For the computation of these forces, we used the IPS method \cite{Wu2005, Wu2009}, a cutoff algorithm based on a statistical description of the images isotropically and periodically distributed in space. 

Let us consider an infinite system obtained when assuming PBC for an elementary
cubic cell with edges of length $L$.
Under the hypothesis that the system is homogeneous on a length scale 
$R_{cut}$,  an effective interaction potential for the molecules
contained in the elementary cell is given by
\begin{equation}
U_{\mathrm{IPS}}(r)=
\begin{cases}
& U(r)+\phi_{\mathrm{IPS}}(r)\ , \qquad  r\leq R_{cut}\\
& 0 \ , \qquad  r>R_{cut}\\
\end{cases}
\end{equation}
where $\phi_{\mathrm{IPS}}(r)$ is a correction to the potential energy which takes in account
the interactions of a single particle with the isotropically distributed images of the system. Such a potential can be written as the sum of two
contributions.
The first one takes into account  the interaction of a test particle 
with the infinite number of images of the source particle along the axes joining the
two particles of the pairwise potential
\begin{equation}
\phi_{\mathrm{IPS}axial}(r)=\xi\sum_{m=1}^{\infty}[U(2mR_{cut}-r)+U(2mR_{cut}+r)]\ .
\end{equation}
The other contribution is given by the average of the potential over all the
images isotropically distributed around the region delimited by the cutoff
radius
\begin{equation}
\phi_{\mathrm{IPS}random}(r)=\sum_{m=1}^{\infty}\left[n(m)-2\xi\right]\phi_{shell}(r,m)\ .
\end{equation}
where $n(m)=24m^2+2$ is the number of images in a shell around the source 
particle with radii $[(2m-1) R_{cut};(2m+1) R_{cut}]$ in 3D-space, and
\begin{equation}
\begin{split}
\phi_{\mathrm{IPS}shell}(r,m)=&\dfrac{1}{2}\int_{0}^{\pi} \bigr[ U\bigr(r^2+(2mR_{cut})^2+\\
&-4mR_{cut}\bigr)\cos\theta\bigr]^{1/2}\sin\theta\mathrm{d}\theta
\end{split}
\end{equation}
is an average of the contribution to the potential of the images which 
are not along the axes joining the test and source particles.
It has to be noticed that for some potentials the series appearing in the
equations above do not converge: for this reason a different reference level
is chosen for the potential energy, i.e.
\begin{equation}
\begin{split}
\phi_{\mathrm{IPS}}(r)=&[\phi_{\mathrm{IPS}axial}(r)-\phi_{\mathrm{IPS}axial}(0)]+\\
&+[\phi_{\mathrm{IPS}random}(r)-\phi_{\mathrm{IPS}random}(0)] \ .
\end{split}
\end{equation}
Finally, the parameter $\xi$ is set to a value such that the force due to the effective potential vanishes 
on the cut off radius:
\begin{equation}
\dfrac{\partial }{\partial r}\left[U(r)+\phi_{\mathrm{IPS}}(r)\right]\Biggr|_{r=R_{cut}} = 0 \ .
\end{equation}
The IPS potential allows to control the introduction of spurious effects due to the presence
of infinite replicas of the system in a way that generalises Ewald's sums also to potentials other than
the electrostatic one.

Assuming that the system is homogeneous on a length scale $R_{c}$, we can define an effective pairwise IPS potential $U^{IPS}=U^{IPS}(|\boldsymbol r_{i,j}|,R_c)$ which takes into account the sum of pair interactions within the local region around a particle
\begin{equation}
U^{IPS}(|\boldsymbol{r}_{i,j}|,R_{c}) = \begin{cases} U(|\boldsymbol{r}_{i,j}|)+\phi(|\boldsymbol{r}_{i,j}|,R_{c})\;, & |\boldsymbol{r}_{i,j}|\leq R_{c} \\ \\ 0\;, & |\boldsymbol{r}_{i,j}|>  R_{c} \end{cases}
\end{equation}
where $\phi(|\boldsymbol{r}_{i,j}|,R_{c})$ is a correction to the potential obtained by computing the total contribution of the interactions with the particle images beyond the cutoff radius $R_{c}$ \cite{Wu2005, Wu2009}. For the Coulomb potential of Eq.\eqref{Coulomb}, we obtained an analytical expression for the IPS correction $\phi_{\text{Coul}}(\boldsymbol{r}_{i,j},R_{c})$. For computational reasons this has been approximated  by a polynomial of degree seven in $x=|\boldsymbol{r}_{i,j}|/R_c$ with $x$ in the interval $(0;1]$:
\begin{equation}
\label{Coulomb_IPS}
\begin{split}
\phi_{\text{Coul}}(x)= &-9.13636 \times 10^{-7}+0.000100298 x +\\
&+ 0.298588 x^2  +0.0151595 x^3 +\\
&+ 0.00881283 x^4 +  0.10849 x^5+ \\
 &-0.0930264 x^6 + 0.0482434 x^7
\end{split}
\end{equation} 
We have chosen $R_c=L/2$ under the hypothesis that on this scale the system is homogeneous.

\subsection{Long-time diffusion coefficient}
We aim at assessing the experimental detectability of long-range interactions between biomolecules using quantities accessible by means of standard experimental techniques. A meaningful approach to this issue is the study of transport properties. For this reason, in our simulations we chose the long-time diffusion coefficient $D$  as the main observable of the system described by Eqs.\eqref{lang2.}. This coefficient is defined, consistently with Einstein's relation \cite{Allen1989}, as:
\begin{equation}
D=\lim_{t\rightarrow +\infty}\dfrac{\langle |\Delta\boldsymbol{r}_i(t)|^2\rangle}{6t} \;,
\label{Dsdef2}
\end{equation}
$\Delta \mathbf{r}_i(t)=\boldsymbol{r}_i(t)-\boldsymbol{r}_i(0)$ being the total displacement of a particle in space and $\langle  a_{i}\rangle=1/N\sum_{i=1}^N a_{i}$ the average over the particle set. We remark that in our system the displacements $\Delta \boldsymbol{r}_i (t)$ are not mutually independent due to the interaction potential $U(|\boldsymbol{r}_i-\boldsymbol{r}_j|)$ in Eqs.\eqref{lang2.}, which establishes a coupling between different particles; in that case, the average over particles index concerns correlated stochastic variables. Nevertheless, as our system is non-linear with more than three degrees of freedom, it is expected to be chaotic \cite{Hirsch2004} so that, in this case, the statistical independence of particle motions is recovered. Moreover, when a chaotic diffusion gives $\langle |\Delta\boldsymbol{r}_i(t)|^2\rangle\propto t$ (which is the case of the models considered in the present work), the diffusion coefficient $D$ is readily computed through a linear regression of $\langle|\Delta\boldsymbol{r}_i (t)|^2\rangle$ expressed as a function of time. In what follows we refer to $\langle |\Delta \boldsymbol{r}_i (t)|^2\rangle$ as Mean Square Displacement (MSD).

\subsection{Simulation Parameters} \label{subsec:SimPar}
Molecular Dynamics simulations were performed considering a solution with $N_A=500$ and $N_B=50$ representing respectively Lysozyme molecules and AF488 molecules. This choice seemed to be a good compromise between the need of a large $N_B$ for a good statistics, a sufficiently large ratio $N_A/N_B$ and the request of a not too high computation time. 
\begin{figure}[h] 
\includegraphics[scale=0.37,keepaspectratio=true]{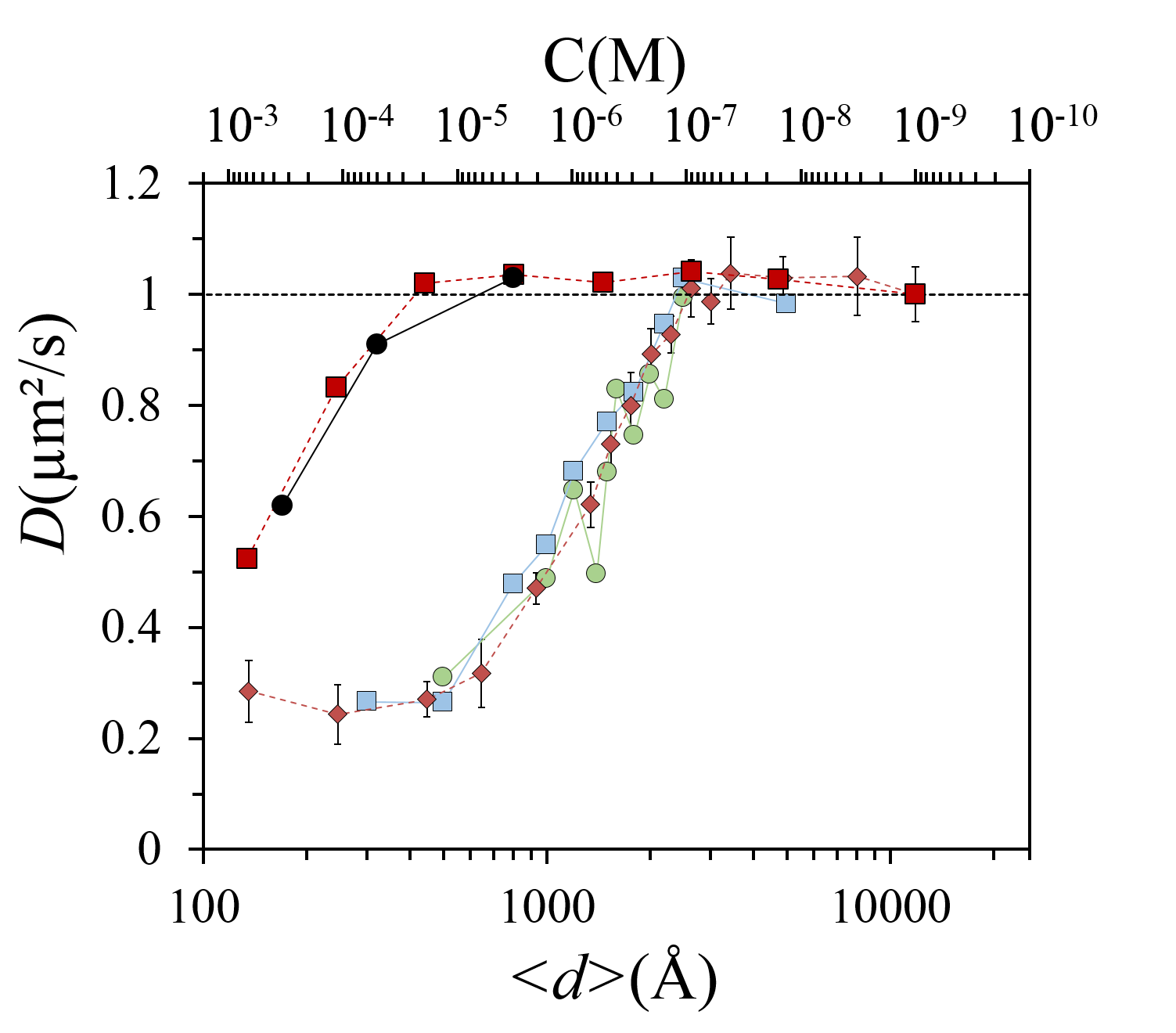}
\caption{ (Color online) Comparison among experimental results and numerical simulations for non-screened Coulomb potential.. Semilog plot of the normalized diffusion coefficients $D/D_0$: experimental values (red diamonds), $N_A=200$ and $N_B=20$ (light green circles), $N_A=500$ and $N_B=50$ (light blue squares). Red squares and full black circles correspond to experimental and numerical results, respectively, in the case of a Debye-H\"uckel potential with a Debye length $\lambda_D=27 \mathring{A}$.}
\label{exp-vs-numl} 
\end{figure}
The $A$-type particles, which represent Lysozyme molecules, have a radius $R_A=2 \times 10^{-3}\,\mu$m and a net electric charge expressed in elementary charge units $Z_A=+10$. The $B$-type particles, which represent AF488 molecules, have a radius $R_B=0.5\times 10^{-3}\,\mu$m and a net electric charge expressed in elementary charge units $Z_B=-2$. 

The medium where diffusion takes place represents an aqueous solution, so that the viscosity $\eta=8.90 \times 10^{-4}\text{Pa}\cdot\text{s}^{-1}$ is the water viscosity at $T\simeq 300 K$. The relative dielectric permittivity has been taken to be $\varepsilon=80$ as for pure water.

The time step has been chosen to $\Delta t=5\times 10^{-4}\,\mu$s: this choice can allow important compenetration among particles  in MDS and does not permit a correct description of excluded volume effects, i.e. the drift due to stochastic forces on dye molecules in a single time steps is comparable with $R_B$. Nevertheless the effects we are interested in concern the diffusive behaviour of dyes on larger length and time scales and we considered acceptable such compenetrations.
The number of time steps was fixed such that the dynamics was simulated for $t_{max}=5000 \mu$s so that for the dye particles $\sqrt{6\,D_{0} t_{max}}\simeq 4\,\mu\text{m}\gg\langle d \rangle$, where $D_{0}$ is the brownian self-diffusion coefficient of the dye molecules, for any considered case. These 5 ms-long simulations are considered adequate  because this time interval is much longer than  the relevant dynamical processes (like trapping and free diffusion) which occur in the real system on much shorter time scales, as estimated in Section \ref{acf-data}. 
In \autoref{exp-vs-numl} the outcomes of the above described numerical simulations are compared to the experimental results obtained for the same quantity: the diffusion coefficient $D$, normalized with respect to 
its Brownian value $D_0$, as a function of the average interparticle distance $\langle d\rangle$. The numerical outcomes for $D$ quantify the diffusion of the small particles that model the dye molecules.  We observe that the choice $N_A=500$ and $N_B=50$ yields a less noisy pattern with respect to the choice $N_A=200$ and $N_B=20$, what is of course sound.  
In \autoref{exp-vs-numl} we have also reported the outcomes of numerical simulations performed with $N_A=500$ and $N_B=50$ and replacing the Coulombic potential with a screened one, that is,  the Debye-H\"uckel potential. We used a Debye length $\lambda_D=27 \mathring{A}$. In so doing a direct comparison can be made with the experimental outcomes obtained with 50 mM of NaCl in solution. Also in this case the agreement is very good.
Even though the number of particles considered in  our numerical simulations is very small with respect to the actual number of molecules in laboratory experiments the agreement among numerical and experimental results is excellent. This is not surprising because it is a common situation in standard Molecular Dynamics simulations. The birth and success of Molecular Dynamics was just due to the possibility of obtaining good values of macroscopic observables out of numerical simulations performed with a few hundreds of particles \cite{rahman}.

\section{concluding remarks}
The work reported in the present paper concludes a feasibility survey aimed at assessing the adequacy of diffusion studies to detect long-range intermolecular forces, hence - among them -  electrodynamic intermolecular interactions in a future adequate experimental setup. 
The two preceding works of Refs. \cite{pre1,pre2} dealt with this problem from the theoretical and numerical sides, respectively. The present work contains a leap forward in what it provides  an experimental assessment of the adequacy of Fluorescence Correlation Spectroscopy to detect intermolecular long range interactions.
Even though our ultimate goal is to detect long range \textit{electrodynamic} intermolecular interactions,  for the time being we have tested this technique against a system where long range interactions are built-in, that is, a solution of oppositely charged molecules interacting through non-screened electrostatic interactions. As a matter of fact, we have found that FCS is certainly appropriate to detect intermolecular interactions in dilute systems, that is, when the solvated molecules interact at large distances, in the present study up to $2500  \mathring{\text{A}}$ approximately.
Furthermore, the excellent quantitative agreement between the experimental outcomes and the corresponding numerical simulations has a twofold relevance. From the one side it confirms that the observed phenomenology, namely, the sudden bending of the diffusion coefficient when the average intermolecular distance is lowered below a critical value,  as well as its pattern as a function of the intermolecular distance, are actually due to the electrostatic interaction among the solvated molecules. From the other side this  validates the numerical algorithm and approximations adopted, suggesting that this numerical scheme can be safely applied to interpret the readouts of experiments where electrodynamic interactions will be possibly excited.

\begin{acknowledgments}
The authors acknowledge the financial support of the Future and Emerging Technologies (FET) Program within the Seventh Framework Program (FP7) for Research of the European Commission, under the FET-Proactive TOPDRIM Grant No. FP7-ICT-318121. This work was also supported by the Minist\`ere de l'Enseignement Sup\'erieur et de la Recherche (ANR-10-INBS-04 France BioImaging and ANR-11-LABX-0054 Labex INFORM to D.M.), Aix-Marseille Universit\'e (ANR-11-IDEX-0001-02 A*MIDEX to D.M.), and institutional funding from the Centre National de la Recherche Scientifique and the Intitut National de la Sant\'e et de la Recherche M\'edicale. The authors are pleased to acknowledge also the financial support of the ''Fondation Princesse Grace'',  Principaut\'e de Monaco (to PF), and a financial support of the Region PACA (to MP, SJ \& PF).
\end{acknowledgments}

\section*{appendix} \label{Ap}

\subsection{Quenching of the dye}
The fluorescence intensity of AF488 is known to be influenced by a quenching effect that the protein exerts on the fluorophore via four aminoacids: Trytophan and Tyrosine  (strong quenchers) and Histidine and Methionine (weaker quenchers) \cite{ChenWebb2007, ChenWebb2010}. These effects are attributed to photoinduced electron transfer (PET) occurring  when the two molecules are in close contact, thus due to short-range interactions ($ < 2 \mathring{A} $) \cite{Lakowicz}.This van der Waals contact takes place on time scales which are not resolved by our FCS apparatus.
When AF488 binds to the Lysozyme the consequent conformational rearrangement influences the dye diffusion time and it also changes the dye fluorescence quantum yield \cite{Melo2014}.
The non specific binding of AF488 on the protein surface is supported by the observed fluorescence quenching factor of about 1.6 between free AF488 and AF488 bound to Lysozyme, as shown in the table reported in \autoref{Tableau}. This value is compatible with steady state fluorescence measures reported in the literature \cite{Melo2014} where the same quenching factor has been found equal to $1.9$.
 
\subsection{Checking possible crowding effects}
\begin{figure}[ht] \centering
\includegraphics [width=75mm]{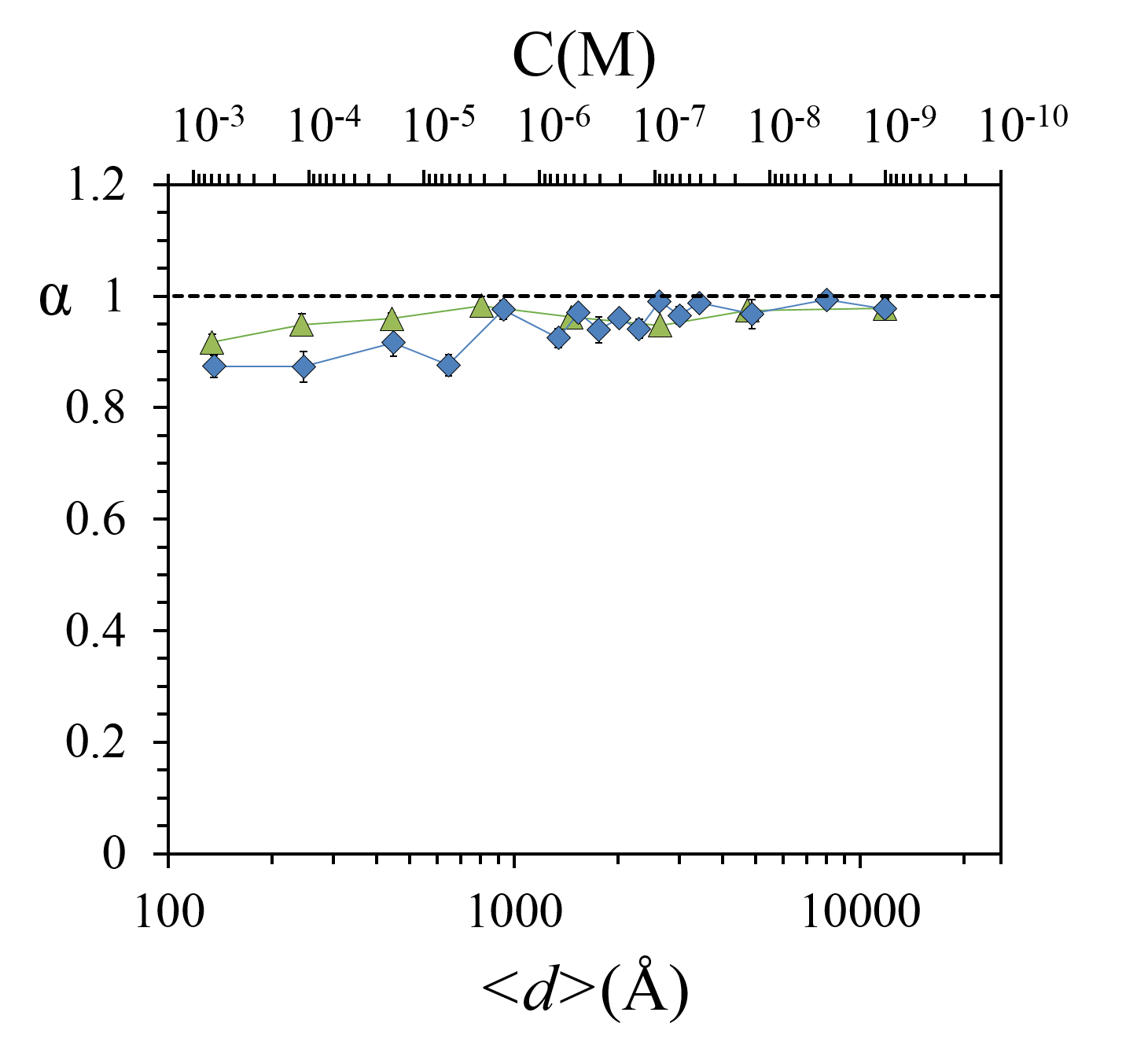}
\caption{(Color online) Check of Brownian versus anomalous diffusion. The $\alpha$ values reported here refer to 0 mM of NaCl in solution ( blue diamonds), and 100mM (green triangles). $\alpha=1$ corresponds to Brownian diffusion, $\alpha < 0.6 - 0.7$ can be attributed to anomalous diffusion.  }
\label{ADalpha} 
\end{figure}

Even though a-priori we do not expect any relevant role played by molecular crowding at our low working concentrations, we have also fitted our ACFs by means of the following analytic expression 
\begin{equation}\label{ACFalpha}
G(\tau)=1+\dfrac{1}{N} \, \dfrac{1+n_T\,\exp{\left(-\dfrac{\tau}{\tau_T}\right)} }{\left[1+\left(\dfrac{\tau}{\tau_D}\right)^\alpha\right]\sqrt{1+s^2\left(\dfrac{\tau}{\tau_D} \right)^\alpha }}\ .
\end{equation}
where the anomalous exponent $\alpha$ becomes a free parameter in the fitting. Using FCS, it has been shown that anomalous diffusion, which corresponds to a mean square displacement of the molecules proportional to $t^\alpha$ with $\alpha$  smaller than 1, is  an indication of the degree of molecular crowding. 
\autoref{ADalpha}  clearly indicates that crowding effects are negligible because there is no evidence of anomalous diffusion which is commonly assumed when $\alpha< 0.6 - 0.7$ \cite{ellis}. 

\subsection{Comparison between FCS and FCCS outcomes}
\begin{figure}[ht] \centering
\includegraphics [scale=0.6,keepaspectratio=true]{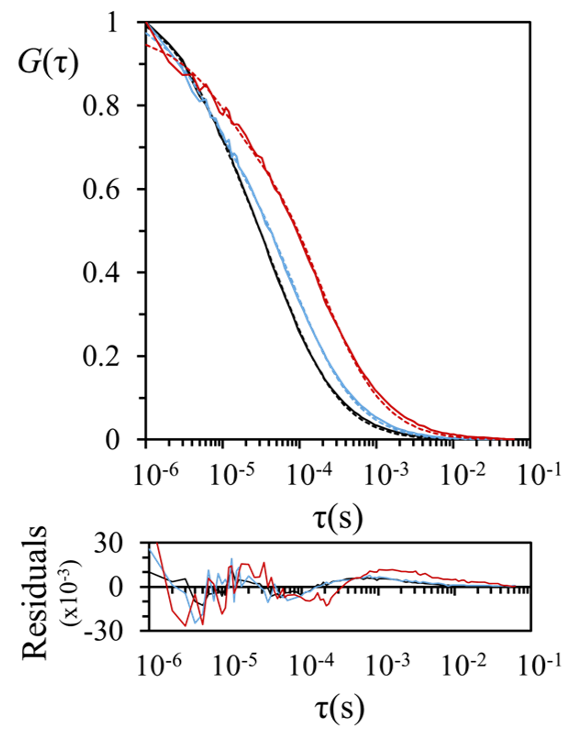}
\caption{(Color online) FCCS measurements. Semilog plot of the normalized autocorrelation function of fluorescence fluctuations, defined in Eq.\protect{\eqref{ACF}}, obtained at $\langle d\rangle =240 \mathring{\text{A}}$ (red line), at $\langle d\rangle =920 \mathring{\text{A}}$ (blue line), and at $\langle d\rangle =4200  \mathring{\text{A}}$ (black line). Working temperature 20$^o$C.}
\label{ACFplotbis} 
\end{figure}

In \autoref{ACFplotbis} some typical outcomes of the FCCS measurements are displayed. These are cross-correlation functions of the fluorescence intensity fluctuations 
$\delta F_1(t)$ and $ \delta F_2(t)$ measured by two independent photo-detectors to eliminate afterpulsing artefacts. The cross-correlation functions are defined by

\begin{equation}
G(\tau) = \dfrac{\langle \delta F_1(t) \delta F_2(t+\tau)\rangle}{\langle F_1(t)\rangle\langle F_2(t)\rangle}
\label{CCFexp}
\end{equation}
(for graphical reasons the normalized versions are displayed). The reported cross-correlation functions correspond to the same concentrations (average intermolecular distance) of the autocorrelation functions reported in \autoref{ACFplot}. Some difference in their shape is observed at very short times (where afterpulsing artefacts are expected), but this does not significantly affect the diffusion time, apart from the fact that the FCS measurements were performed at 30$^o$C, whereas the FCCS measurements were performed at 20$^o$C (because of a technical constraint of our FCCS equipment).

\begin{figure}[h!] \centering
\includegraphics [width=90mm]{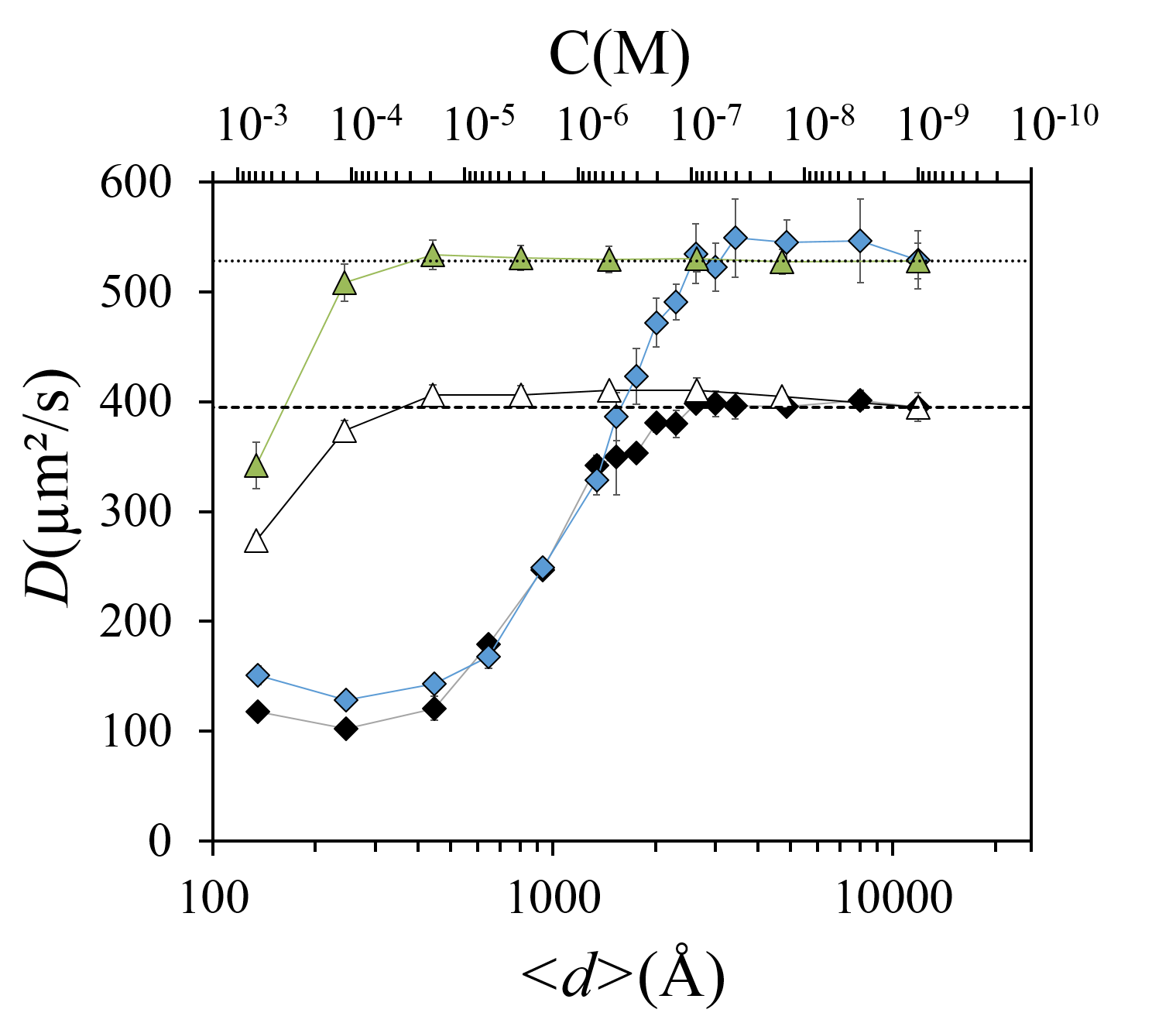}
\caption{(Color online) Comparison between the FCS  and the FCCS measurements. Semilog plot of the diffusion coefficients $D$ of AF488 as a function of the distance in  
$\mathring{\text{A}}$ - and of the concentration of the solution in Moles - between proteins and dyes.The FCS results have been obtained at 30$^o$C,  0 mM  (blue diamonds), and 100mM (green triangles), of NaCl in solution. The FCCS results have been obtained at 20$^o$C,  0 mM  (black diamonds), and 100mM (white triangles), of NaCl in solution.  }
\label{FCS-FCCS} 
\end{figure}

\begin{figure}[h] \centering
\includegraphics [width=170mm]{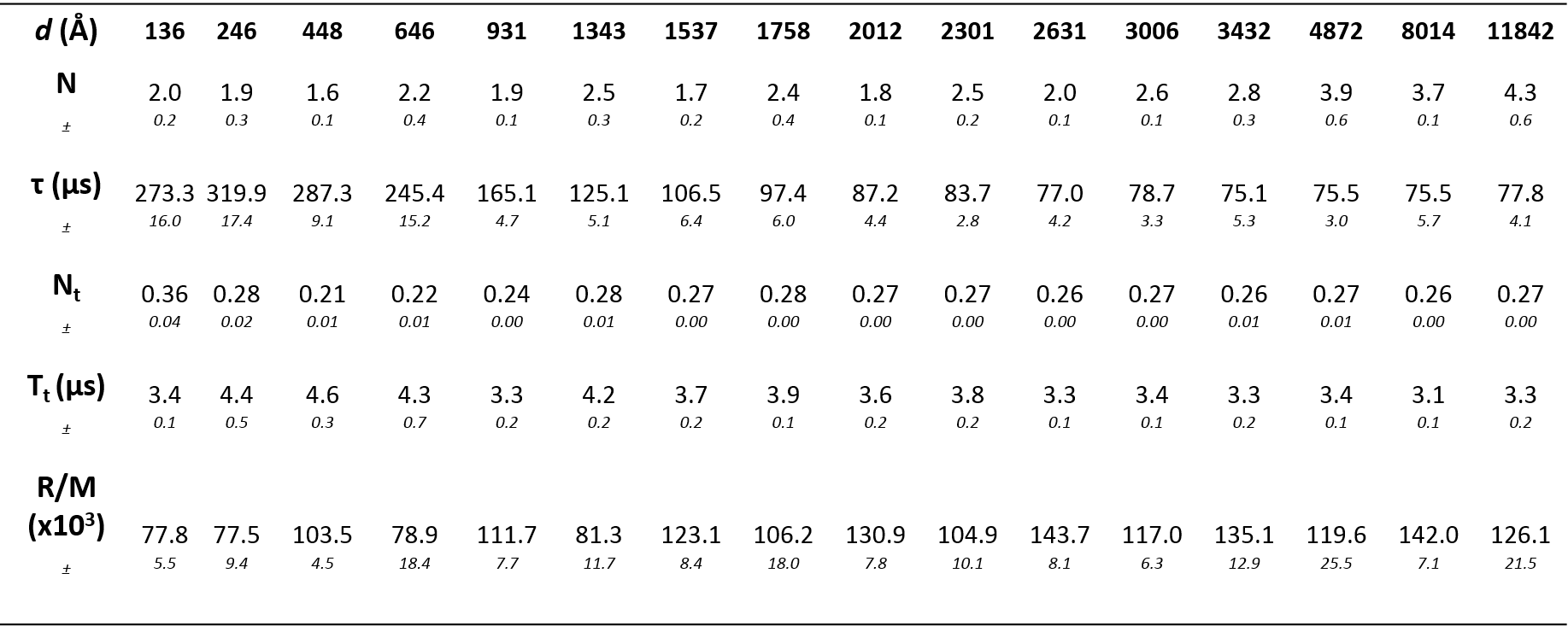}
\caption{Table reporting the fitted values of the parameters for solutions of variable concentrations of Lysozyme and 1nM of AF488.  Solutions containing 0mM of NaCl (blue diamonds visible in \autoref{D-D0_D_Rh}, \autoref{dye-Lys}, \autoref{D-D0_NaCl} and \autoref{FCS-FCCS}). Each value is the result of the averaging of the data recorded during 20 measurements on four different samples and for three independent experiments. The fitted parameters are: the number of molecules (N), the diffusion time in microseconds ($\tau$), the number of molecules in the triplet state ($N_{t}$), the time spent in the triplet state in microseconds ($T_{t}$) and the emission rate (number of photons emitted per second) per molecule (R/M). All these data  are displayed according to the average intermolecular distance between the molecules given in Angstroms ($\langle d\rangle$).}
\label{Tableau} 
\end{figure}

A comparison between the non normalised values of the diffusion coefficient obtained with FCS and FCCS is provided in \autoref{FCS-FCCS}. The experimental points lying on the horizontal lines correspond to Brownian diffusion of AF488 molecules; the discrepancy between these values is explained by the temperature difference, in fact, by inverting Equation \eqref{stokes} to get $D=k_BT/(6\pi R_H\eta)$, with $R_H(AF488)\simeq 5.2\mathring{A}$, $\eta (20^oC) = 10^{-3}$ Pa s, and $\eta(30^oC)= 0.797\times 10^{-3}$ Pa s,  we obtain $D_{20^o} = 410 \mu$m$^2$ s$^{-1}$ and $D_{30^o} = 532 \mu$m$^2$ s$^{-1}$,  respectively. Whence the ratio $D_{30^o}/D_{20^o} = 1.3$ to be compared with the value $1.31$ of the fraction of the Brownian diffusion coefficients reported in \autoref{FCS-FCCS}.


In \autoref{FCS-FCCS} we also observe a discrepancy in the transition value of $\langle d\rangle$, which appears smaller in the FCCS case. Again, this is a temperature dependent effect the physics of which is qualitatively understood by inspecting Eqs. \eqref{lang2.}. In fact, by lowering the temperature of the solution the viscosity of water $\eta$ increases, and the coefficient $1/\gamma =1/(6\pi R_H\eta)$ results in a weakening of the  Coulombic interactions; then also the strength of thermal noise is weakened but only through a $\sqrt{1/\gamma}$ factor so that the net effect is a reduction of the strength of the Coulombic interactions.
\bigskip
\clearpage


\begin{thebibliography}{99}
\bibitem{nota-concentr} We proceeded by diluting a drop of a 100 nM watery solution of the AF488 dye into a larger quantity of Lysozyme solution so as to keep constant the dye concentration in the final mixture. If $N_0/V_0$ is the initial concentration of the dye, and if $N_p/V_p$ is the initial concentration of the protein, then $C_{AF488}=N_0/(V_0+V_p)$ and $C_{Lys}=N_p/(V_0+V_p)$ so that Equation \protect{\eqref{Av-dist}} trivially follows.
\bibitem{nota} If there are $M$ different species of non-interacting diffusing particles  in the observation volume, the overall fluorescence autocorrelation is 
$$
G(\tau)=\dfrac{1}{ N }  \sum_{i=1}^{M}
\dfrac{f_i [1 + \ n_{T}\exp{(-\tau /\tau_{T}}) ]   }{\left(1+\dfrac{\tau}{\tau_{D,i}}\right) 
 \sqrt{1+s^{2}\dfrac{\tau}{\tau_{D,i}}}}
$$
where $N$ is the total number of molecules in the confocal volume. The $M$ populations with diffusion times $ \tau_{D,i} $ must  satisfy the condition:  $ \sum_{i=1} f_i=1 $.
\end{thebibliography}
\end{document}